\documentclass[12pt]{amsart}

\usepackage{amsmath,amsthm,amssymb,amsfonts}
\usepackage{mathrsfs}
\usepackage{graphicx} 
\usepackage[font=small]{caption}
\usepackage[all,cmtip]{xy}

\makeatletter
\DeclareRobustCommand{\rvdots}{%
  \vbox{
    \baselineskip4\p@\lineskiplimit\z@
    \kern-\p@
    \hbox{.}\hbox{.}\hbox{.}
  }}
\makeatother

\begin{document}

\title{Predictive analysis of microarray data}

\author[Marques]{Paulo C. Marques F.}
\address{Instituto de Matem\'atica e Estat\'istica da Universidade de S\~ao Paulo}
\email{pmarques@ime.usp.br}
\author[Pereira]{Carlos A. de B. Pereira}
\date{December 9, 2013}

\begin{abstract}
Microarray gene expression data are analyzed by means of a Bayesian nonparametric model, with emphasis on prediction of future observables, yielding a method for selection of differentially expressed genes and the corresponding classifier.
\end{abstract}


\maketitle

\section{Introduction}

DNA microarrays are devices used to determine the expression (activity level) of a set of genes contained in a tissue sample. Briefly, they consist of small arrays of thousands of probes on which surfaces are deposited many copies of single stranded DNAs sequences corresponding to specific genes, or pieces of genes. Reverse transcription of messenger RNAs extracted from the tissue produces a solution of DNAs whose sequences are complementary to those found on the microarray probes. This solution is colored and put into contact with the microarray surface. Sequences present in the solution hybridize with their complementary pairs on the microarray probes. Subsequent illumination of the microarray surface provides an image in which the intensity of each probe spot is related to the corresponding amount of messenger RNAs present in the tissue. Digital processing of this image outputs for each probe a positive number which measures the relative expression of the corresponding genes (see \cite{friend} and references therein for a detailed description of microarray technology).

Data from a typical microarray experiment consist of positive numbers representing the expression levels of the genes associated with the microarray probes for a group of individuals. Because the convoluted nature of the numeric values describing the expression levels makes it difficult to commit to a specific family of probability distributions in their modeling, our proposal is to approach this problem by means of a Bayesian nonparametric analysis. The emphasis placed by De Finetti \cite{definetti} on prediction guides us, in the sense that both products of our analysis, a subset of differentially expressed genes and the corresponding classifier, are derived from probabilities of events related to values of future observables, with (unobservable) parameters playing only a subsidiary role.

\section{Microarray data model}

Our microarray data consist of the expression levels of $p$ gene probes for $m$ case patients that have been diagnosed with a certain disease or show some physiological alteration, and $n$ healthy control individuals. The expression level of the $j$-th microarray probe for the $i$-th case patient is denoted by $X^j_i$. Similarly, expression levels for controls are denoted by $Y^j_i$. The expression levels of the $p$ gene probes for the $i$-th case patient are abbreviated by $X_i=(X^1_i,\dots,X^p_i)$. For controls, we define similarly $Y_i=(Y^1_i,\dots,Y^p_i)$.

The graph below depicts the microarray data model. Absence of an arrow connecting two random objects means that they are conditionally independent given their parents. In this graph, the orphan vertexes are independent Dirichlet processes distributed as $F_j\sim\mathrm{DP}(c_j,F_0^j)$ and $G_j\sim\mathrm{DP}(d_j,G_0^j)$, for gene probes $j=1,\dots,p$. Necessary Dirichlet process properties and notations are collected in the first appendix.

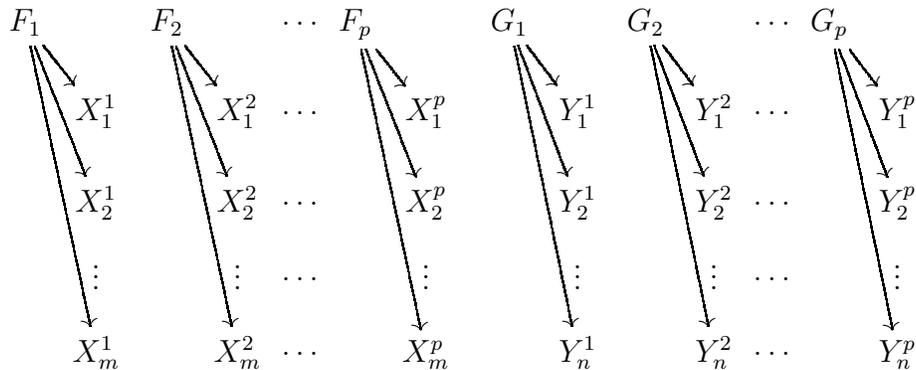
\[
  \SelectTips{cm}{}
  \vcenter{\xymatrixcolsep{0.14cm}\xymatrixrowsep{0.5cm}\xymatrix{
   F_1 \ar@//[dr] \ar@//[ddr] \ar@//[ddddr] & *{} & F_2 \ar@//[dr] \ar@//[ddr] \ar@//[ddddr] & *{} & *{\dots} & F_p \ar@//[dr] \ar@//[ddr] \ar@//[ddddr] & *{} & *{} & G_1 \ar@//[dr] \ar@//[ddr] \ar@//[ddddr] & *{} & G_2 \ar@//[dr] \ar@//[ddr] \ar@//[ddddr] & *{} & *{\dots} & G_p \ar@//[dr] \ar@//[ddr] \ar@//[ddddr] & *{} \\
   *{} & X^1_1 & *{} & X^2_1 & *{\dots} & *{} & X^p_1  & *{} & *{} & Y^1_1 & *{} & Y^2_1 & *{\dots} & *{} & Y^p_1 \\
   *{} & X^1_2 & *{} & X^2_2 & *{\dots} & *{} & X^p_2  & *{} & *{} & Y^1_2 & *{} & Y^2_2 & *{\dots} & *{} & Y^p_2 \\
   *{} & *{\rvdots} & *{} & *{\rvdots} & *{\dots} & *{} & *{\rvdots}  & *{} & *{} & *{\rvdots} & *{} & *{\rvdots} & *{\dots} & *{} & *{\rvdots} \\
   *{} & X^1_m & *{} & X^2_m & *{\dots} & *{} & X^p_m  & *{} & *{} & Y^1_n & *{} & Y^2_n & *{\dots} & *{} & Y^p_n
  }}
\]

\section{Predictive selection of differentially expressed genes}

Suppose that we have microarray gene expression data for $m$ case patients and $n$ healthy control individuals. Following the notations introduced in the previous section and using the convention that upper case letters represent random variables and small case letters their realizations, we denote this data by $\{x_i\}_{i=1}^m$ and $\{y_i\}_{i=1}^n$. Our first goal is to use the information contained in this data to establish which genes are expected to be less or more active for a future case patient, making these differentially expressed genes a subset of disease markers.

For each gene probe, the relative expression of the corresponding gene can be determined by our posterior opinion that for a future case patient the expression level of this probe will be smaller than the expression level of the same probe for a future healthy control individual, given all the information contained in the data. This posterior opinion is quantified by the posterior predictive probabilities
$$
  q_j=\mathrm{Pr}\!\left( X_{m+1}^j\leq Y_{n+1}^j\mid \{X_i=x_i\}_{i=1}^m, \{Y_i=y_i\}_{i=1}^n\right) \, ,
$$
for gene probes $j=1,\dots p$. Using the nonparametric data model of the previous section, these posterior predictive probabilities can be computed using the results given in the first appendix.

The values of these probabilities determine an ascending ranking of relative expression of the gene probes. We denote by $X_i^{(j)}$ the expression level of the gene probe occupying the $j$-th position in this ranking for the $i$-th case patient. To refer to the gene probes at the end of the ranking, we use the notation $X_i^{(-j)} = X_i^{(p-j+1)}$. Similar notations, $Y_i^{(j)}$ and $Y_i^{(-j)}$, are used for healthy control individuals.

The criterion for the choice of the subset of differentially expressed genes is to select from this ranking the first $k$ (down regulated) gene probes and the last $k$ (up regulated) ones, for some integer $k\geq 1$. In the last section we show how $k$ can be selected by cross validation.

\section{Predictive classification}

With the subset of $2k$ differentially expressed gene probes obtained in the previous section, we construct a classification rule that allows us to pick microarray data for a new individual and classify him as unhealthy or healthy. Let us denote case patients and healthy controls together as
$$
  (Z_1,\dots,Z_{m+n}) = (X_1,\dots,X_m,Y_1,\dots,Y_n) \, .
$$

Defining the statistic $T(Z_i) = \prod_{j=1}^k Z_i^{(-j)}/Z_i^{(j)}$, which is an increasing function of the expression level of the up regulated gene probes and a decreasing function of the down regulated ones, we expect case patients to exhibit values for this statistic that are larger than the corresponding values for healthy controls. 

The classification rule is based on this one dimensional statistic and takes into account the different sample sizes of the two groups, cases and controls. Given a new individual for which we have microarray data $z_{m+n+1}$, the rule is to classify him as healthy when $T(z_{m+n+1})$ is less than the critical value $t^*$ for which
$$
  \mathrm{Pr}\!\left(T(Z_{m+n+1})\leq t^*\mid \{Z_i=z_i\}_{i=1}^{m+n}\right) = \frac{n}{m+n} \, ;
$$
otherwise, we classify him as unhealthy. The critical value $t^*$ is computed using the results given in the first appendix.

\section{Example}

The publicly accessible Gene Expression Omnibus (GEO) database \cite{geo} provides microarray data from a study \cite{pan} of peripheral circulating B cells for $m=39$ smoking and $n=40$ non-smoking healthy american white women.

We proceed the analysis of this dataset using the results of the previous sections and considering the case of weak prior information about the distribution of the expression levels of the gene probes, which means that within the nonparametric model we compute all the desired probabilities taking the limit to zero of the concentration parameters of the Dirichlet processes.

Table \ref{table:tdcv} shows the identifiers for the $k=4$ pairs of down and up regulated gene probes computed for this dataset. This table also presents a leave one out cross validated study of the sensitivity and specificity of the predictive classifier for this dataset using the $k=4$ pairs of down and up regulated gene probes. For this dataset, $k=4$ is the smallest number of pairs of up and down regulated gene probes that gives us the best balance between cross validated fractions of false negatives and false positives. Computer code in the Perl language \cite{wall} is presented in the second appendix.

\begin{table}
\def~{\hphantom{0}}
\caption{Gene probes and cross validation}
\begin{tabular}{cccrrr}
Down regulated & Up regulated & $\qquad\qquad$ & & Unhealthy & Healthy \\
EBP & GPR15 & $\qquad\qquad$ & Case & 100\% & 0\% \\
EIF4B & DDX3X & $\qquad\qquad$ & Control & 2.5\% & 97.5\% \\
H3F3AP4 & CBFB & $\qquad\qquad$ & & & \\
MFSD11 & SCAF11 & $\qquad\qquad$ & & & \\
\end{tabular}
\label{table:tdcv}
\end{table}

\section*{Acknowledgments}

This paper is dedicated to Paulo Cilas Marques \textit{in memoriam}. We thank Professor Luiz Eug\^enio Barbosa de Oliveira for his critical reading of the manuscript. Work partially supported by CAPES.

\section*{Appendix 1: The Dirichlet process}\label{app:dp}

We are concerned with the representation of our uncertainties about some observable properties assuming values in a sampling space $\mathscr{X}$, with sigma-field $\mathscr{A}$, by means of a probability measure defined over an underlying measurable space $(\Omega,\mathscr{F})$. The probability of an event $B\in\mathscr{F}$ is denoted by $\mathrm{Pr}(B)$. 

The map $Q:\mathscr{A}\times\Omega\to[0,1]$ is a random probability measure over $(\mathscr{X},\mathscr{A}) $ if $Q(\,\cdot\,,\omega)$ is a probability measure over this measurable space for every $\omega\in\Omega$, and $Q(A)=Q(A,\cdot\,)$ is a random variable for each $A\in\mathscr{A}$.

Ferguson \cite{ferguson} defined a random probability measure $Q$ as follows. Let $\alpha$ be a finite nonzero measure over $(\mathscr{X},\mathscr{A})$ and specify that for each $\mathscr{A}$-measurable partition $\{A_1,\dots,A_k\}$ of $\mathscr{X}$ the random vector $$(Q(A_1),\dots,Q(A_k))$$ has the usual Dirichlet distribution with parameters $(\alpha(A_1),\dots,\alpha(A_k))$. One such $Q$ is denominated a Dirichlet process with base measure $\alpha$.

Ferguson proved that this definition entails the following facts. First, $Q$ is a properly defined random process in the sense of Kolmogorov's consistency theorem \cite{schervish}. Second, the expectation of $Q$ has the simple expression $\mathrm{E}[Q(A)]=\alpha(A)/\alpha(\mathscr{X})$, for each $A\in\mathscr{A}$. Third, if measurable observables $X_i:\Omega\to\mathscr{X}$ are conditionally independent and identically distributed, given $Q$, with $\mathrm{Pr}\!\left(X_i\in A\mid Q\right)=Q(A)$ almost surely, for $i=1,\dots,m$, then \textit{a posteriori} $Q$ is again a Dirichlet process with base measure $\beta$ defined almost surely by $\beta(A)=\alpha(A)+\sum_{i=1}^m I_A(X_i)$, for each $A\in\mathscr{A}$.

If we add a new observable $X_{m+1}$ to the just described conditional model, its posterior predictive probability is
\begin{align*}
  \mathrm{Pr}\!\left(X_{m+1}\in A\mid \{X_i\}_{i=1}^m\right) &= \mathrm{E}\!\left[  \mathrm{Pr}\!\left(X_{m+1}\in A\mid Q,\{X_i\}_{i=1}^m\right) \mid \{X_i\}_{i=1}^m\right] \\
&=\mathrm{E}\!\left[\mathrm{Pr}(X_{m+1}\in A\mid Q) \mid \{X_i\}_{i=1}^m\right] \\
&=\mathrm{E}\!\left[Q(A)\mid \{X_i\}_{i=1}^m\right] ,
\end{align*}
almost surely, for every $A\in\mathscr{A}$, in which the second equality follows from the conditional independence of the observables.

For microarray data the sampling space can be taken as the real line with Borel sigma-field. If $Q$ is a Dirichlet process with base measure $\alpha$, it is convenient to work with the random distribution function defined by $F(t,\omega)=Q((-\infty,t],\omega)$. We abbreviate $F(t)=F(t,\cdot\,)$. Defining $c=\alpha(\mathbb{R})$ and $F_0(t)=\alpha(-\infty,t]/\alpha(\mathbb{R})$, we denote the distribution of the random distribution function by $F\sim\mathrm{DP}(c,F_0)$. Since $\beta(\mathbb{R})=c+n$, the posterior expectation of $F$ is almost surely 
$$
  \hat{F}_{0,m}(t) = \mathrm{E}\!\left[F(t)\mid\{X_i\}_{i=1}^m\right] = \frac{c}{c+m} F_0(t) + \frac{m}{c+m} \hat{F}_m(t) \, ,
$$
in which $\hat{F}_m(t)=(1/m)\sum_{i=1}^m I_{[X_i,\infty)}(t)$ is the empirical distribution function. This gives us an interpretation of the base measure of the Dirichlet process. The total measure $c$ works as a concentration parameter: for fixed sample size $m$, if we make $c\downarrow 0$, the posterior expectation reduces to the empirical distribution function. Also, this expression of $\hat{F}_{0,m}$ shows that prior information contained in $F_0$ is washed out when, for fixed $c$, we let $m\to\infty$.

Finally, suppose that we have a second sample: let $Y_1,\dots,Y_n,Y_{n+1}$ be conditionally independent and identically distributed, given $G$, each one of them having conditional distribution $G$, and $G\sim\mathrm{DP}(d,G_0)$ is independent of $F$. The posterior expectation is almost surely
$$
  \hat{G}_{0,n}(t) = \mathrm{E}\!\left[G(t)\mid\{Y_i\}_{i=1}^n\right] = \frac{d}{d+n} G_0(t) + \frac{n}{d+n} \hat{G}_n(t) \, ,
$$
in which $\hat{G}_n(t)=(1/n)\sum_{i=1}^n I_{[Y_i,\infty)}(t)$. If $U$ and $V$ are independent random variables with distribution functions $F_U$ and $F_V$, respectively, a simple computation shows that $\mathrm{Pr}(U\leq V)=\int_{-\infty}^\infty F_U(t)\,dF_V(t)$. Therefore, since $X_{m+1}$ and $Y_{n+1}$ are conditionally independent, given $\{X_i\}_{i=1}^m$ and $\{Y_i\}_{i=1}^n$, and almost surely
$$
    X_{m+1}\mid\{X_i\}_{i=1}^m\sim \hat{F}_{0,m} \, , \quad Y_{n+1}\mid\{Y_i\}_{i=1}^n\sim \hat{G}_{0,n} \, ,
$$
it follows that almost surely
$$
  \mathrm{Pr}\!\left(X_{m+1} \leq Y_{n+1}\mid \{X_i\}_{i=1}^m,\{Y_i\}_{i=1}^n\right) = \int_{-\infty}^\infty \hat{F}_{0,m}(t)\,d\hat{G}_{0,n}(t) \, .
$$
If we let $c,d\downarrow 0$, this conditional probability reduces to 
$$\frac{1}{mn}\sum_{i=1}^m\sum_{j=1}^n I_{[X_i,\infty)}(Y_j) \, .$$

\section*{Appendix 2: Computer code}\label{app:perl}

{\SMALL
\begin{verbatim}
#!/usr/bin/perl

# predictive.pl - <pmarques@ime.usp.br>

use strict;
use warnings;

my $k = 4;

print "\nSelecting k = $k pairs of Down / Up regulated gene probes.\n\n";

my @cases = (42..80);
my @controls = (2..41);
my @all = (@cases, @controls);

my $ua;

open(DATA, "./GDS3713.soft") or die $!;
while (<DATA>) {
    next unless $. >= 118 && $. <= 22400;
    chomp;
    push @$ua, [ split(/\t/, $_) ];
}
close(DATA);

my @pr;
foreach my $probe (@$ua) {
    push @pr, pr_next_case_leq_next_control($probe, \@cases, \@controls);
}

my @ranking = sort { $pr[$b] <=> $pr[$a] } (0 .. @$ua - 1);

print "Down regulated | Up regulated\n";
print "---------------+--------------\n";
for (my $i = 0; $i < $k; $i++) {
    printf("%-14s | %-13s\n", 
           $ua->[$ranking[$i]]->[1], $ua->[$ranking[-($i + 1)]]->[1]);
    
}
print "\n";

printf("Critical t = %.4f\n\n", 
       critical_t($ua, \@ranking, $k, scalar @controls, \@all));

print "Cross validated sensitivity and specificity.\n\n";

my ($case_unhealthy, $case_healthy, 
    $control_unhealthy, $control_healthy) = (0, 0, 0, 0);

foreach my $group (\@cases, \@controls) {
    foreach my $off (@$group) {
        my $s = T_statistic($ua, \@ranking, $k, $off);
        my @T;
        foreach my $individual (@all) {
            next if $individual == $off;
            push @T, T_statistic($ua, \@ranking, $k, $individual);
        }
        @T = sort { $a <=> $b } @T;
        if ($group == \@cases) {
            if ($s <= $T[@controls - 1]) { $case_healthy++ }
            else                         { $case_unhealthy++ }
        } else {
            if ($s <= $T[@controls - 2]) { $control_healthy++ }
            else                         { $control_unhealthy++ }
        }
    }
}

print "-" x 29, "\n         Unhealthy | Healthy\n";
print "-------------------+---------\n";
printf("    Case    %.4f |  %.4f\n", 
       $case_unhealthy / @cases, $case_healthy / @cases);
print "-" x 29, "\n";
printf(" Control    %.4f |  %.4f\n", 
       $control_unhealthy / @controls, $control_healthy / @controls);
print "-" x 29, "\n";

exit 1;

sub pr_next_case_leq_next_control {
    my ($probe, $cases, $controls) = @_;
    my $leq = 0;
    foreach my $case (@$cases) {
        foreach my $control (@$controls) {
            $leq++ if $probe->[$case] <= $probe->[$control];
        }
    }
    return $leq / (@$cases * @$controls);
}

sub T_statistic {
    my ($ua, $ranking, $k, $individual) = @_;
    my $t = 1;
    for (my $i = 0; $i < $k; $i++) {
        $t *= $ua->[$ranking->[-($i + 1)]]->[$individual];
        $t /= $ua->[$ranking->[$i]]->[$individual];
    }
    return $t;
}

sub critical_t {
    my ($ua, $ranking, $k, $n, $all) = @_;
    my @T;
    foreach my $individual (@$all) {
        push @T, T_statistic($ua, $ranking, $k, $individual);
    }
    @T = sort { $a <=> $b } @T;
    return $T[$n - 1];
}
\end{verbatim}
}

\end{document}